\documentclass[conference]{IEEEtran} 
\usepackage[colorlinks]{hyperref}           
\usepackage{tabularx,booktabs}
\usepackage{hhline}
\usepackage{fancyhdr}
\usepackage{alphalph}
\usepackage[font=small,labelfont=bf]{caption}
\usepackage[flushleft]{threeparttable}
\ifCLASSINFOpdf
   \usepackage[pdftex]{graphicx}
\else
\fi

\usepackage{amsmath}
\usepackage{array}

\usepackage[dvipsnames]{xcolor}
\newcommand\myshade{85}
\colorlet{mylinkcolor}{blue}
\colorlet{mycitecolor}{green}
\colorlet{myurlcolor}{green}

\hypersetup{
  linkcolor  = mylinkcolor!\myshade!black,
  citecolor  = mycitecolor!\myshade!black,
  urlcolor   = myurlcolor!\myshade!black,
  colorlinks = true,
}

\usepackage{enumitem}
\setlist{leftmargin=4mm}
\usepackage{multirow}
\usepackage{url}

\begin{document}

\title{Optimizing Smart Grid Aggregators and Measuring Degree of Privacy in a Distributed Trust Based Anonymous Aggregation System}


\author{\IEEEauthorblockN{Mohammad Saidur Rahman}
\IEEEauthorblockA{Global Cybersecurity Institute \\
Rochester Institute of Technology\\
Rochester, NY, USA.\\
Email: \texttt{saidur.rahman@mail.rit.edu}}
}

\maketitle
\thispagestyle{plain}
\pagestyle{plain}
\begin{abstract}
A smart grid is an advanced method for supplying electricity to the consumers alleviating the limitations of the existing system. It causes frequent meter reading transmission from the end-user to the supplier. This frequent data transmission poses privacy risks. Several works have been proposed to solve this problem but cannot ensure privacy at the optimal level. This work is based on a distributed trust-based data aggregation system leveraging a secret sharing mechanism. In this work, we show that {\em three aggregators} are enough for ensuring consumer's privacy in a distributed trust-based system. We leverage the idea of anonymity in our research and show that neither an active attacker nor a passive attacker can breach consumer's privacy. We show proof of our concept mathematically and in a cryptographic game based mechanism. We name our new proposed system \emph{``Distributed Trust Based Anonymous System (DTBAS)''}.
\end{abstract}


\IEEEpeerreviewmaketitle

\section{Introduction}
The present scenario of electricity supply, meter reading, and consumer service is not as advanced as it could be. To make the supply of electricity more advanced, smart grid is proposed. Unlike current electric grid system in which meter reading is accomplished bi-weekly or monthly basis, smart grid suggests frequent data transmission (i.e. 15-minutes of interval) from the meters to the utility. The objectives behind this proposition are to provide better service to the consumer, solve problem rapidly, managing the supply and use of the smart grid more efficiently \cite{lipton2017smart}.\par 

There are several benefits of introducing smart grid. It can ensure sustainability and reduce carbon dioxide \cite{bohli2010privacy}. As smart meters will provide the electricity usage details frequently, it can motivate the users to reduce their consumption and minimize their utility cost. On the supplier side, it can help electric supplier to introduce dynamic pricing mechanisms \cite{wood2003dynamic}. Like all the technical advancements, this advancement also depicts some challenges.\par

As smart meter increases the flow of customer daily electricity usage data precisely to the electricity supplier, it introduces privacy challenge. If the supplier is malicious, or any other party gets those precise personal data, the client's privacy is breached. An attacker can know when the client is home, and she can plan for targeted attack. The clients fall into the risk of targeted marketing too. To cope with these problems, several researches have been published to protect user's privacy by aggregating data based on different cryptography protocols \cite{bos2017privacy} \cite{pan2017efficient}\cite{barletta2015privacy}\cite{biselli2013protection}\cite{garcia2010privacy}.\par

\begin{figure}[!t]
	\centering
    \includegraphics[scale=0.7]{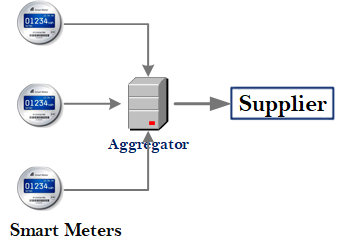}
    \caption{Data Aggregation Model in Smart Grid.}
    \label{fig_steer}
\end{figure}

The core model is to collect the meter reading data in the aggregators and send the aggregated data to the supplier. The purpose is to obscure the direct raw data transmission from the client to the supplier (\textcolor{blue}{Figure-}\ref{fig_steer}).
The existing cryptography based models cannot ensure complete trust in the process of data transmission. The cryptography based protocols are acceptable to data transmission. However, it cannot ensure the protection of complete privacy. The distributed trust is more secure than trust in a single system. Hence, distributing the smart meter data into multiple aggregators may enhance the privacy. But there is no previous work that answered how many aggregators can be enough to distribute the trust. Though some research proposes more than two aggregators \cite{buescher2017two}. However, in the distributed trust mechanisms using secret sharing, it is yet to know the optimized number of aggregators.\par
In this work, we show that three aggregators are enough for a distributed trust based system. And three aggregators are enough to protect user's privacy. In our proposed system, the number of aggregators cannot be less than three and the number of users cannot be less than three too. A supplier can increase the number of aggregators if the capacity of the aggregators is overloaded. We leverage the concept of anonymity in a system \cite{diaz2002towards}.\par
In anonymity literature, there has been different mechanisms to ensure data anonymity and connection anonymity \cite{diaz2002towards}. Data anonymity means the actual data that are being transferred and the connection anonymity means to make the sender of the data anonymous. Interestingly, in our proposed distributed trust based anonymous system, we can ensure both data and connection anonymity. We can measure the degree of anonymity based on the power of an attacker. For example, how much data an attacker can gather and how much an attacker can know from the collected traffic or data. When the degree of anonymity is one, all the users of our system being the originator of the sent data have equal probability. Claudia et. al. \cite{diaz2002towards} used Shanon's definition of entropy \cite{shannon2001mathematical} to quantify the degree of anonymity. We take the same method to measure the degree of anonymity in our proposed system. As anonymity is measured considering the power of an attacker. We consider both active and passive attacker. We will give a detail description of those two attack scenarios in section-4.1. We provide a mathematical proof of our proposed system in section-5.\par
We also provide proof of our proposed system in a cryptographic game based privacy metric proposed by Bohli et. al. \cite{bohli2010privacy}. Niklas et. al. \cite{buescher2017two} used this approach in their work to measure the privacy. In this approach, the privacy is measured by the disadvantage of an attacker to distinguish between two users. The game is based on two parties: the adversary and the challenger. The success of the adversary depends on successfully identifying the user from a set of users that the challenger provides. We give the detail description in \textcolor{blue}{Section-} \ref{cryptographic_game}.


\section{Related Work} \label{lit_review}
The breach of privacy because of the frequent meter data transmission is not desirable by any client. It can reveal information about a client's family, electricity usage pattern, and also specific-time information about a client which are certainly scary \cite{greveler2012multimedia}. For example, Alice is watching HBO at 10.00PM. Some research has proposed client's privacy protection through anonymous data communication from the smart meter to the supplier.\par

Pan et al. proposes an aggregation scheme eliminating the need for a TTP and dividing the users into various groups \cite{pan2017efficient}. They leveraged the chinese remainder theorem and paillier key encryption (PKE) scheme to design their system. However, they still do not answer the question of how many aggregator we need. Engel and Eibl propose an approach called Wavelet-based multi-resolution that is based on combining multiple resolutions and direct user control for smart meter (SM) \cite{engel2017wavelet}. In this system, aggregators collect encrypted real time SM readings from individual users relying on distribution operators (Wavelet Encryption). Silva et al. tries to solve the limitations to ensure SM privacy using Intel SGX SDK \cite{silva2017security}. They conclude that Intel SGX can provide simple and general solution for SM privacy problem. However, they tend to make the communication stronger in different cryptographic mechanism but fail to provide solution of the quest how many aggregators can be best for the communication purpose.\par

It has been also a challenge to measure the privacy with a standard privacy metrics. Buescher et al. \cite{buescher2017two} measure the privacy based on Bohli et al.'s \cite{bohli2010privacy} proposed approach that is based on cryptographic game. Though their privacy metric is widely used, to make their system work, it requires a lot of users (i.e. 4,50,000). Hence, we are not adopting that metric.\par

\subsection{Anonymity in the Smart Grid}
Many researches have conducted to study the privacy of the smart meters by anonymizing consumers consumption data. A study conducted by Efthymiou and Kalogridis \cite{efthymiou2010smart}, they proposed a system that anonymizes metering data that sent by smart meters which are utility consumption data or operational data. They applied two different identifiers, low frequency identifier for sending utility bills or operational purposes, and high frequency identifier for specific locations data. High frequency identifier is authenticating by third party escrow service to make it difficult to associate it with specific SM or costumer.\par
But they require the presence of a trusted third party (TTP) in the system. However,  TTP is not a sustainable solution as it can increase the system complexity \cite{pan2017efficient} and be a malicious entity too. Hence, we aim to alleviate the need to a trusted third part.  Instead aggregation based mechanism can be more secure and enhance the privacy without being dependent on the TTP. In this mechanism, the smart meters communicate encrypted data to each other before going for the aggregation \cite{garcia2010privacy}.\par

Another study done by Ford  et al. \cite{ford2017secure}, they proposed a protocol which is stored all data in Trusted Third Party (TTP). (TTP) acts as anonymous identifier that responsible for analyzing and computing usage data, and utility provider request billing information and some final result from (TTP), which means the data is divided between (TTP) and utility provider. So, no one of them has a full record of the consumers usage data. This schema is centralized on (TTP), and it is vulnerable to single point failure. Moreover, if the (TTP) gets compromised or they change the amount of data that they provide it to the utility provider, consumer’s privacy gets violated.

\subsection{Secret Sharing}
There are many ways to prevent secret to be discovered, one is to divide the secret into multiple shares which are should be collected together to get the secret again. There are a bunch of researchers conducted studies in this field. A study done by Shamir \cite{shamir1979share} , it showing a system based on polynomial interpolation by dividing the secret into a number pieces in a way that it can be easy reconstruct from any share pieces, but with uncompleted shares don’t give any information about the secret. Another study done by Asmuth and Bloom \cite{asmuth1983modular}, which is showing a scheme similar to Shamir’s scheme for reconstruction the secret, it relies on the Chinese Remainder Theorem. 
While some studies are investigated different aspects of secret sharing by studying the degree of security to protect it against malicious attempts. Feldman \cite{feldman1987practical} has done an investigation about a protocol in verifiable secret sharing (VSS), which is a cryptography tool for distributed systems such as smart meters. It’s basically guarantees that any share can be verified in which secret is belonging to, against any compromise of corruption by a malicious.\par
There is a study mentioned the privacy preserving of smart meters using secret sharing scheme. Rottondi et al. \cite{rottondi2012security} proposed a framework that is responsible for protecting consumers information by providing different levels of aggregations without revealing for any party individual information by applying Shamir’s scheme in their framework. Moreover, they proposed an infrastructure for collecting the data for each consumer which is the Privacy Preserving Nodes (PPN). It’s basically relying secret sharing scheme, consumer info is the secret and it needs to divide into shares and each (PPN) carrying a share for a specific secret, then aggregate the shares to the system according to each customer. The full information (the secret) can be retrieved by collecting all shares from (PPN), the (PPN) is acting as the aggregators.



\section{Research Design} \label{research_design}
The intended focus of this project is to propose a distributed trust based anonymous system (DTBAS) for smart grid's data aggregation. In this system, the optimal number of aggregators proposed. Afterwards, we measure the degree of anonymity of our proposed DTBAS. The strength of any proposed system depends of the ability of that system to defend particular attacks. We need to have a well-defined attack model that we are aiming to defend. We are assuming the presence of both active and passive attacker in our attack models. For the purpose of trust based system, we are going to adopt the secret sharing mechanism and 15-minutes interval of time for the transmission of meter reading data. The cryptographic mechanism to transmit data from the smart meters to the aggregators is not the focus of this project. The readings of a single smart meter in 15-minutes time interval will look like (\textcolor{blue}{Table-}\ref{tab:smreading}).
\vspace*{1 cm}
\begin{table}[h!]
\fontsize{10}{10}
    \caption{Smart Meter Readings in Distributed Trust based System in 15-minutes time interval.}
    \label{tab:smreading}
    \begin{tabular}{c c c c c c c }
       \hline
       & & $AG_1$ & $AG_2$ &$AG_3$ & $.....$ & $AG_n$ \\
      \hline \hline

     $15-mins$& $t_1$ & $t_{11}$ & $t_{12}$ & $t_{13}$ & $.....$ & $t_{1n}$\\
     $15-mins$& $t_2$ & $t_{21}$ & $t_{22}$ & $t_{23}$ & $.....$ & $t_{2n}$\\
     $45-mins$& $t_3$ & $t_{31}$ & $t_{32}$ & $t_{33}$ & $.....$ & $t_{3n}$\\
     $.....$& $.....$ & $.....$ & $.....$ & $.....$ & $.....$ & $.....$\\
     $30-days$& $t_m$ & $t_{m1}$ & $t_{m2}$ & $t_{m3}$ & $.....$ & $t_{mn}$\\

       \hline \hline
    \end{tabular}
\end{table}

The reading of a time interval is divided into sub-reading. As mentioned earlier, the secret sharing mechanism is intended to be implemented in each of the sub-reading. \par 
In this section, we provide the attack models, system model and the measurement model of our proposed DTBAS.

\subsection{Attack Model} \label{design_attack_model}
As we are proposing an anonymous system, the degree of anonymity is measured based on the power of an attacker in a particular attack model. This proposed system may not work in different attack models. Hence, a clear concept and definition of the attack model is required. We are using the same definition Diaz et. al. \cite{diaz2002towards} used in their work with little modification to define our active and passive attacker.\par 

In our attack models (i.e. active and passive), the attacker is capable of performing probabilistic attack. She can assign probabilities of being the originator of data in a specific client. This kind of attacks are known as probabilistic attack \cite{raymond2001traffic}. 

\subsubsection{Active Attack} \label{active_attack_model}
An attacker is said to be an active attacker if she can exploit or control at least one clients of a system. In other words, she can see the data that are passing through the system and she can even prevent the client from sending any data to the system. \par 

\begin{figure}[ht]
	\centering
    \includegraphics[scale=0.7]{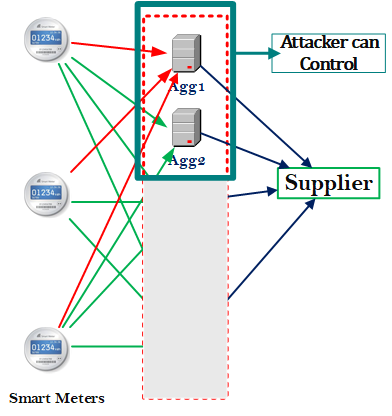}
    \caption{Active Attack Model in DTBAS.}
    \label{active_attacker}
\end{figure}

However, in our anonymous data aggregation system, the active attacker is assumed to have power to control or exploit at least one or at best two aggregators. The attacker can access the information received by those aggregators, but cannot successfully identify what data belongs to which client (\textcolor{blue}{Figure-}\ref{active_attacker}).

\subsubsection{Passive Attack} \label{passive_attack_model}

We are adopting local-global attacker's definition of Diaz et. al. \cite{diaz2002towards} as our passive attacker in our proposed system. This kind of attacker can posses the control of the entire systems.\par 

\begin{figure}[ht]
	\centering
    \includegraphics[scale=0.7]{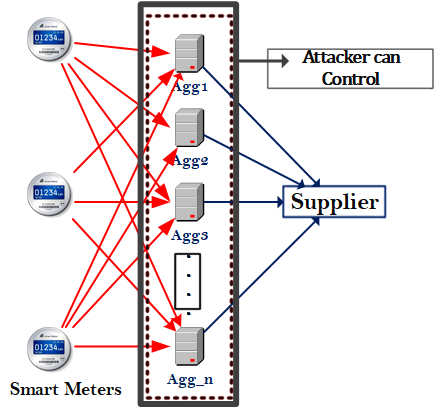}
    \caption{Passive Attack Model in DTBAS.}
    \label{passive_attacker}
\end{figure}

In our anonymous data aggregation system, the passive attacker is assumed to have the power to control the whole aggregation systems. This attacker in this system can be the supplier of the electricity. We can assume that, only the supplier can have the power to access all the aggregators of the proposed system (\textcolor{blue}{Figure-}\ref{passive_attacker}).

\subsection{System Model} \label{design_system_model}
\begin{figure}[ht]
	\centering
    \includegraphics[scale=0.5]{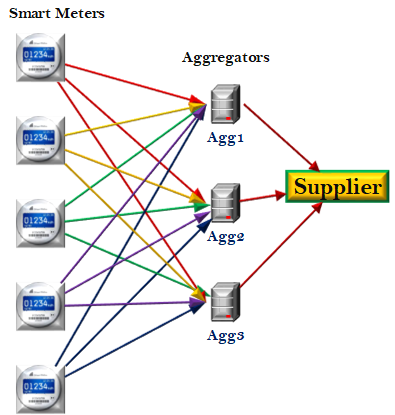}
    \caption{Distributed Trust Based Anonymous Aggregation System (DTBAS).}
    \label{system_model}
\end{figure}

We are proposing a distributed trust based anonymous aggregation system (DTBAS) with only three aggregators. We aim to achieve
anonymity in our system through the split of smart meter's data into three aggregators (\textcolor{blue}{Figure-}\ref{system_model}). The descriptions of the two parties of our system are given as follows.\par

\
\par
\begin{figure}[ht]
	\centering
    \includegraphics[scale=0.5]{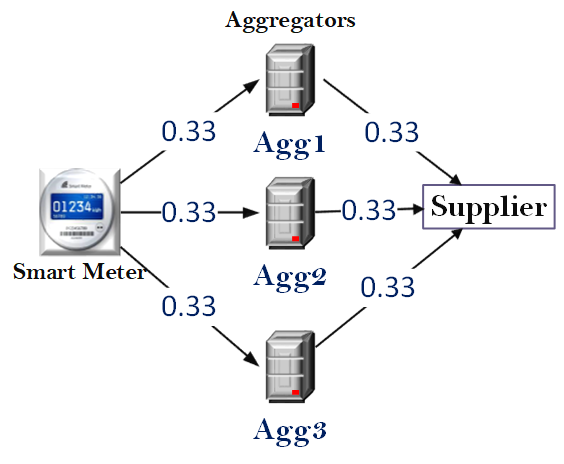}
    \caption{Data Transmission from Smart Meter to Aggregators.}
    \label{system}
\end{figure}

\textbf{\textit{Senders (Smart Meters)}}: Smart meters are the sender of data to the aggregators. Smart meters are the entity of our proposed system which anonymity we aim to protect. Smart meters send data in 15-minutes time interval. Each smart meter divides the reading data into three equal part and then send different part to different aggregators. That means each aggregator will get 1/3 of the whole meter reading data (\textcolor{blue}{Figure-}\ref{system}) of a single smart meter which is also the major concept of distributed trust based mechanism.  By splitting whole data into three splits, we are achieving anonymity along with distributed trust.

\
\par
\begin{figure}[ht]
	\centering
    \includegraphics[scale=0.5]{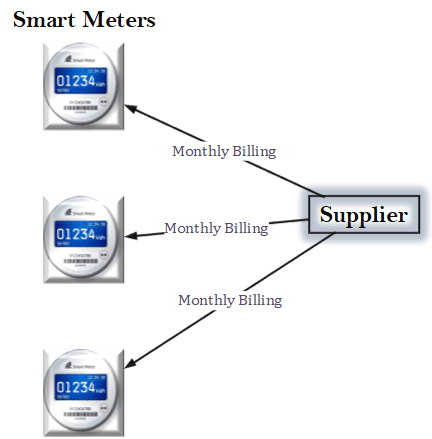}
    \caption{Flow of Billing Data from the Supplier to the Smart Meters.}
    \label{billing}
\end{figure}

\textbf{\textit{Receivers (Aggregators)}}: In our proposed system, the receiver are the aggregators that receive data from the smart meters. The communication mechanism of the aggregators is one-way. By one-way, we mean that it does not respond or send back any data into the smart meters. The supplier directly sends billing data to the smart meters (\textcolor{blue}{Figure-}\ref{billing}). The aim is to defend any correlation attack that an attacker can perform from the aggregated data and the billing data. It might be possible for an attacker to deanonymize the users by the correlation attack. Our objective is to lower the severity of information that an attacker might use to perform any kind of attack.

\subsection{Measurement Model} \label{design_measurement_model}
As we are proposing an anonymous aggregation system, the definition of the anonymity needs to well-defined. Given the splits of data from the smart meters to the aggregators in our system, the anonymity set will be the  total number of users. In the system, the total number of nodes will be the multiplied result of the total number of users and the number of splits.

\[ \text{Let,}\ n=\text{Total Number of Users}. \]
\[  m=\text{Number of Splits/Number of Aggregators}. \]
\[ \text{Hence,}\ Anonymity \ Set = \ n \]
\[ \text{Total Number of Nodes in the System} = \ m*n \]

In this work, we aim to protect the anonymity of the smart meters $ n $ which are the users. In our system, we considers the users as honest. The definition of honest users is that the attacker cannot exploit the behavior of the smart meters (i.e. the smart meters cannot be malicious).\par 

It is intuitive that the anonymity set must consists of more than two users. Otherwise, an attacker can assign the probability of 50\% to each users.  At the same time, the number of splits or aggregators must be more than two. The splits are the portion of data, an attacker will get 50\% of the data. The attacker will have greater advantage to analyze half of the data to deanonymize an user. That is why we proposed three aggregators in our system.
\[ The\ number\ of\ Users > 2 \]
\[ The\ number\ of\ Aggregators > 2 \]

\subsubsection{Measuring Degree of Anonymity} The highest degree of anonymity is calculated when an attacker can find out all the users of an anonymity set and can assign probabilities to each of the users. At the same time, the probabilities of the users being the senders of the data are equal.\par

In our proposed system, the degree of anonymity does not depend on the size of the anonymity set $n$. Rather, we calculate the anonymity based on the information it can gain and assign probabilities to the users as being the senders of the information. The information an attacker can gain depends on the number of splits or aggregators in our system. For example, if the splits are three, an attacker can gain 33\% of information by exploiting an aggregator, and if the splits are four, an attacker can gain 25\% of information by exploiting an aggregator. We are taking the same measurement formulas Diaz et. al. \cite{diaz2002towards} used in their work.
The entropy $H(ES)$ of the system is calculated as \textcolor{blue}{Equation-}\ref{eqn_entropy}:

\begin{equation}
	\label{eqn_entropy}
    H(ES) = - {\sum^{s}_{i=1}p_i\;log_2 (p_i)}
\end{equation}

\[ Here, \ ES\ =\ Entropy \ of \ the \ system. \]
\[ s\ = \ Number\ of\ Splits. \]

The maximum entropy of our proposed system $H(MaxES)$ is measured as \textcolor{blue}{Equation-}\ref{eqn_max_entropy}:
\begin{equation}
	\label{eqn_max_entropy}
    H(MaxES) = {log_2 (s)}
\end{equation}

The degree of anonymity $d_a$ of our system is measured as \textcolor{blue}{Equation-}\ref{eqn_anonymity}:
\begin{equation}
	\label{eqn_anonymity}
    d_a = 1 - \frac{H(MaxES) - H(ES)}{H(MaxES)}
\end{equation}

All the users have the equal probability of being the senders of data if and only if $ d_a = 1$.

\section{Proof of Concept} \label{proof}
\subsection{Computing Degree of Anonymity} \label{proof_measuring_anonymity}
We calculated the degree of anonymity with both equal probability, $P_i$ (\textcolor{blue}{Table-}\ref{tab:table1}) and variable probability (\textcolor{blue}{Table-}\ref{tab:table2}). In equal probability, all the splits have the same probability. In other words, the aggregator contains equal amount of information. If an attacker can exploit a particular aggregator, she cannot gain more than that in equal probability. We can see from \textcolor{blue}{Table-}\ref{tab:table1} that aggregators equal to two or more are giving us the degree of anonymity $d_a = 1$. But we cannot choose two aggregators as we mentioned in our system model.
\vspace*{1 cm}
\begin{table}[h!]
  \begin{center}
    \caption{Degree of Anonymity with \textbf{Equal} Probability.}
    \label{tab:table1}
    \begin{tabular}{c c c c c} 
    \hline
      Number of  & \multirow{2}{*}{$P_i$} & \multirow{2}{*}{{\bf $H(ES)$}} & \multirow{2}{*}{$H(MaxES)$} & \multirow{2}{*}{$ d_a$} \\
       Aggregator&  & &  &\\
      \hline \hline
      1 & 1.00&0.00 & 0.00 & None\\
      \hline
      2 & 0.5 & -1.00 & 1.00 & 1.0\\
       & 0.5 &  &  & \\
      \hline
       & \textcolor{blue}{0.33}&  &  & \\
       \textcolor{blue}{3} & \textcolor{blue}{0.33}& \textcolor{blue}{-1.58} & \textcolor{blue}{1.58} & \textcolor{blue}{1.0}\\
         & \textcolor{blue}{0.33}&  &  & \\
      \hline
       & 0.25 &  & & \\
      4 & 0.25 & -2.00 & 2.00 & 1.0\\
      & 0.25 &  & & \\
      & 0.25 &  & & \\
      \hline
       & 0.20 &  &  & \\
       & 0.20 &  &  & \\
      5 & 0.20 & -2.32 & 2.32 & 1.0\\
      & 0.20 &  &  & \\
      & 0.20 &  &  & \\
      \hline \hline
    \end{tabular}
  \end{center}
\end{table}

\begin{table}[h!]
  \begin{center}
    \caption{Degree of Anonymity with \textbf{Variable} Probability.}
    \label{tab:table2}
    \begin{tabular}{c c c c c} 
    \hline
      Number of  & \multirow{2}{*}{$P_i$} & \multirow{2}{*}{{\bf $H(ES)$}} & \multirow{2}{*}{$H(MaxES)$} & \multirow{2}{*}{$ d_a$} \\
       Aggregator&  & &  &\\
      \hline \hline
       & \textcolor{blue}{0.50} &  & \\
     \textcolor{blue}{3} & \textcolor{blue}{0.49} & \textcolor{blue}{- 1.07} & \textcolor{blue}{1.58} & \textcolor{blue}{0.68}\\
       & \textcolor{blue}{0.01} &  & \\
      \hline
       & 0.50 &  & \\
      4 & 0.48 & - 1.14 & 2.00 & 0.57 \\
       & 0.01 &  & \\
       & 0.01 &  & \\
       \hline
       & 0.50 &  & & \\
       & 0.47 & & & \\
      5 & 0.01 & -1.21 & 2.32 & 0.52 \\
       & 0.01 & & & \\
      & 0.01 & & & \\
      \hline \hline
    \end{tabular}
  \end{center}
\end{table}

\begin{table*}[t!]
  \begin{center}
    \caption{Information an Active Attacker can Gain in \textbf{Active Attack} Model.}
    \label{tab:table3}
    \begin{tabular}{c c c c c}
       &\textcolor{black}{ $AG_1$ }& $AG_2$ & $AG_3$ & $SM$ \\
       &  &  &  & $Data$\\ 
      \hline \hline
     $SM_1$ &\textcolor{blue} {$SM_{11}$} & $SM_{12}$ & $SM_{13}$ & \textcolor{red}{$ \sum^{3}_{i=1} SM_{1i}$} \\
     &  &  &  & \\
     \hline
     $SM_2$ &\textcolor{blue} {$SM_{21}$} & $SM_{22}$ & $SM_{23}$ & $ \sum^{3}_{i=1} SM_{2i}$ \\
     &  &  &  & \\
       \hline
     $SM_3$ &\textcolor{blue} {$SM_{31}$} & $SM_{32}$ & $SM_{33}$ & $ \sum^{3}_{i=1} SM_{3i}$ \\
     &  &  &  & \\
       \hline
     $SM_4$ &\textcolor{blue} {$SM_{41}$} & $SM_{42}$ & $SM_{43}$ & $ \sum^{3}_{i=1} SM_{4i}$ \\
     &  &  &  & \\
       \hline
     $.....$ &\textcolor{blue} {$.....$} & $.....$ & $.....$ & $ .....$ \\
     &  &  &  & \\
       \hline
     $SM_n$ &\textcolor{blue} {$SM_{n1}$} & $SM_{n2}$ & $SM_{n3}$ & $ \sum^{3}_{i=1} SM_{ni}$ \\
     &  &  &  & \\
       \hline
     $AG$  & & &  &  \\
     $Data$ &\textcolor{red} {$\sum^{n}_{i=1} SM_{n1}$} & $\sum^{n}_{i=1} SM_{n2}$ & $\sum^{n}_{i=1} SM_{n3}$ & \\
     &  &  &  & \\

       \hline \hline
    \end{tabular}
  \end{center}
\end{table*}

\begin{table}[h!]
  \begin{center}
  	\vspace*{1 cm}
    \caption{Probability of an User being the Originator of the data Decreases as the Number of Users Increases.}
    \label{tab:table4}
    \begin{tabular}{c c}
    	\hline
       \textbf{ $Number\ of\ SM$ }& \textbf{$Probability\ (P_{user})$}  \\
       \hline \hline
       2 & 0.50 \\
       3 & 0.33 \\
       4 & 0.25 \\
       5 & 0.20 \\
       6 & 0.17 \\
       ..... & ..... \\
       n & $1/n$ \\
       \hline \hline
    \end{tabular}
  \end{center}
\end{table}

In the variable probability, $P_i$ (\textcolor{blue}{Table-}\ref{tab:table2}), we are assuming that an attacker can assign random guess probability at least to one aggregator (split) and then we assign probability in decremental manner (i.e. 50\%, 49\%, 0.01\%). The reasons behind this method of assigning probability is that, we are giving the attacker maximum power to know about the particular split. A strong global level passive adversary (\textcolor{blue}{Figure-}\ref{passive_attacker}) who can see everything can assign probabilities in this way. We can see from \textcolor{blue}{Table-}\ref{tab:table2} that the degree of anonymity $d_a = 0.68$ which is the highest with three aggregators. The more we increase the number of aggregators, the attacker gains more advantage. The lower the degree of anonymity, the higher an attacker gain advantage.\par
Observing the both cases to measure the degree of anonymity, we are proposing that three aggregators are enough for our proposed system. As in our active attack model, an attacker cannot simultaneously exploit more than one aggregator. Realistically she cannot even assign these high probability like we are assigning. We are doing that to increase the power of an attacker in our system and show that with this high probability the attacker might fail.

\subsection{Defended Active Attack} \label{proof_active_attack_model}
As defined in our active model in \textcolor{blue}{Section-}\ref{active_attack_model} can exploit or in control an aggregator and she can get the data that is coming to this aggregator. The attacker can obtain the accumulated blue marked information in \textcolor{blue}{Table-}\ref{tab:table3}. If she accumulates, she can see the blue marked accumulated data \textcolor{blue}{$\sum^{n}_{i=1} SM_{n1}$}. To deanonymize the $smart meter 1$, the attacker needs the accumulated data \textcolor{blue}{$\sum^{n}_{i=1} SM_{1i}$}. We can see from \textcolor{blue}{Equation-}\ref{notequal} that the information an attacker can gain is not equal to the information she needs to deanonymize the smart meter. Hence, we can say from this mathematical operation that our defined active attack can be defended by our proposed system.\\
\begin{equation}
	\label{notequal}
	\sum^{n}_{i=1}SM_{n1} \neq \sum^{n}_{i=1}SM_{1i}
\end{equation}

Even if the active attacker can assign high probability in a particular smart meter (maximum of random guess in our model), she cannot gain the whole information because of the distributed trust based aggregation system model. In addition the number of users will be large number. That is why the seeming advantage of an attacker is always lower than the actual advantage.

Realistically, the probability of a smart meter $(SM)$ being the originator of a message will get decreased along with the increase of the smart meters (\textcolor{blue}{Table-}\ref{tab:table4}).

\subsection{Defended Passive Attack} \label{proof_passive_attack_model}
As we have mentioned in \textcolor{blue}{Section-}{\ref{passive_attack_model}} that the passive attacker have the power to control the whole aggregation systems. This attacker in this system can be the supplier of the electricity.\par
Intuitively, it is frightening that the authority is being thee attacker. As the data is encrypted with complex cryptographic mechanisms, she has to go through a long computational process to decrypt the data (\textcolor{blue}{Table-}\ref{tab:table3}). An attacker will need a long time to decrypt the data and map to a specific smart meter. Deanonymization of an user will not be fruitful after a long period to perform an attack as the usage pattern may not remain same after a long time. 

\subsection{Cryptographic Game} \label{cryptographic_game}
Cryptographic game based mechanism is used in the work of Niklas et. al. \cite{buescher2017two}. They leveraged the approached of Bohli et. al. \cite{bohli2010privacy} to measure the privacy.\par
In this mechanism, there are two parties: the adversary and the challenger. The success of the adversary depends on successfully identifying the user from a set of users that the challenger provides.\par
Firstly, The adversary chooses two load profiles $lf_1\ and\ lf_2$ and sends those to the challenger. The challenger then take one of those load profiles and mixes with a set of load profiles $ ( \ lf_1\ / \ lf_2\ ,\ l_3\ , \ l_4 \ ..... l_n\ )$. Afterwards, the challenger this mixes to the adversary. If the adversary can distinguish the from the load profile from the mixes, it is his success. Unless otherwise it is failure.\par 
This process goes on for $5000$ times and the probability of success is calculated based on the number of times the adversary can distinguish the load profile from the mixes.\par
However, in our anonymous aggregation system, the users are of equal probability. In addition the data is of equal portion in each aggregators. Hence, it is confusing for the adversary to distinguish the user from the mixes as all will be of equal probability.

\section{Limitations and Future Work} \label{limit_and_future_work}
Though our proposed anonymous aggregation system model is novel, we cannot say this system is practically implementable unless we experiment with real data and simulate th actual results.\par 
Hence, our future work is aimed to experiment this system with real world electricity consumption data. In addition to that, we aim to test in real world network traffic to cross-validate our system.

\section{Conclusion} \label{conclusion}
In this work, we are proposing a distributed trust based anonymous aggregation system (DTBAS) to send data from the smart meters to the supplier. We are proposing that three aggregators are enough for our proposed system. This novel system model will solve the privacy problem of the electricity users. We proof mathematically the effectiveness of our proposed system in two well-defined attack models (i.e. active attack and passive attack). We measured the degree of anonymity based on the advantage an attacker can gain from the information. We also give explanation of the effectiveness of our model against cryptographic game based approach. We aim to solve the limitations of our proposed system by experimenting in real world dataset in future. 


\bibliographystyle{IEEEtran}
\bibliography{SmartGridPrivacy}

\end{document}